\documentclass[pra,a4paper,twocolumn,superscriptaddress,amsmath,amssymb,floatfix,amstex]{revtex4-1}
\usepackage{graphicx,graphics,rotating}
\usepackage{dcolumn}
\usepackage{bm,lipsum}
\usepackage{hyperref}
\usepackage{color}
\usepackage{soul} 
\usepackage{mathbbol}

\newcommand{\ud}{\mathrm{d}}
\newcommand{\ic}{\mathrm{i}}
\newcommand{\re}{\mathrm{ Re }\;}
\newcommand{\im}{\mathrm{ Im }\;}
\newcommand{\tr}{\mathrm{ tr }\;}
 
\newcommand{\D}[1]{\displaystyle}

\def\Cij{J_{ij}}
\def\Gi{h_i}
\def\A{A_{\bm{s},\bm{s'}}(t)}

\begin{document}
\title{Analytical results for the quantum non-Markovianity of spin ensembles undergoing pure dephasing dynamics}

\author{R\'emy~Dubertrand}
\affiliation{Institut f\"ur Theoretische Physik, Universit\"at Regensburg, 93040 Regensburg, Germany}
\affiliation{Institut de Physique Nucl\'eaire, Atomique et de
Spectroscopie, CESAM, Universit\'e de Li\`ege, B\^at.\ B15, B - 4000
Li\`ege, Belgium}
\author{Alexandre~Cesa}
\affiliation{Institut de Physique Nucl\'eaire, Atomique et de
Spectroscopie, CESAM, Universit\'e de Li\`ege, B\^at.\ B15, B - 4000
Li\`ege, Belgium}
\author{John~Martin}
\affiliation{Institut de Physique Nucl\'eaire, Atomique et de
Spectroscopie, CESAM, Universit\'e de Li\`ege, B\^at.\ B15, B - 4000
Li\`ege, Belgium}

\date{\today}%
\begin{abstract}
We study analytically the non-Markovianity of a spin ensemble, with arbitrary number of spins and spin quantum number, undergoing a pure dephasing dynamics. The system is considered as a part of a larger spin ensemble of any geometry with pairwise interactions. We derive exact formulas for the reduced dynamics of the system and for its non-Markovianity as assessed by the witness of Lorenzo \emph{et al.} [Phys.~Rev.~A \textbf{88}, 020102(R) (2013)]. The non-Markovianity is further investigated in the thermodynamic limit when the environment's size goes to infinity. In this limit and for finite-size systems, we find that the Markovian's character of the system's dynamics crucially depends on the range of the interactions. We also show that, when the system and its environment are initially in a product state, the appearance of non-Markovianity is independent of the entanglement generation between the system and its environment.
\end{abstract}

\maketitle

\section{Introduction}

Open quantum systems can display a large variety of dynamical behaviors, including decoherence \cite{joos_decoherence_2003,zurek_decoherence_1991,haroche_entanglement_1998}, thermalization and memory effects. The notion of non-Markovianity, accounting for memory effects, has found applications in many different fields ranging from quantum optics \cite{de_vega_matter-wave_2008}, quantum thermodynamics \cite{bylicka_thermodynamic_2016,pezzutto_implications_2016}, quantum information theory \cite{bylicka_non-markovianity_2015,bellomo_non-markovian_2007} to quantum foundations \cite{matsuzaki_magnetic_2011,chin_quantum_2012,dhar_characterizing_2015,glick_markovian_2017}. Non-Markovianity has also been identified as a key ingredient to achieve specific tasks in the context of quantum heat machines and quantum information processing \cite{gonzalez-tudela_dissipative_2010,malekakhlagh_non-markovian_2016,fang_non-markovian_2017}.
While Markovian dynamics for discrete variable systems is always governed by a master equation of the Gorini-Kossakowski-Sudarshan-Lindblad (GKSL) type, the methods for treating non-Markovian quantum dynamics and their physical interpretation are generally much more complicated, see e.g. \cite{rivas_quantum_2014,breuer_colloquium_2016,de_vega_dynamics_2017,shibata_generalized_1977,wilkie_positivity_2000,Laine2012}. A direct consequence is that non-Markovian master equations are only rarely analytically solvable \cite{Briegel,Maniscalco}. The departure from Markovian dynamics can be quantified through measures of non-Markovianity (see Sec.~\ref{definitions}). Even when the dynamics of the system and its environment is known, evaluating analytically measures of non-Markovianity is often a difficult task, so that up to now only a limited number of analytical results have been obtained \cite{She17,bhattacharya_exact_2017,fischer_correlated_2007}. The aim of this work is to contribute to the analytical treatment of non-Markovianity in the case of spin ensembles undergoing pure dephasing dynamics, with a particular emphasis on the thermodynamic limit of infinitely many spins in the system and/or the environment. Note that non-Markovianity in spin chains has already been studied analytically in \cite{prosen_exact_2010,mahmoudi_non-markovian_2016} and numerically in  \cite{buca_note_2012,cormick_dissipative_2013,smirne_interaction-induced_2013,ribeiro_non-markovian_2015}.

The paper is organized as follows. In Section \ref{definitions}, we present three different measures of non-Markovianity and introduce our main model of a spin ensemble with arbitrary pairwise interaction range and longitudinal external field. In Section \ref{results}, the reduced dynamics and a non-Markovianity witness are evaluated analytically for such an ensemble. The cases of nearest-neighbor and infinite range interactions are discussed in detail, in particular in the limit of an infinite number of spins. A comparison with other measures of non-Markovianity is presented. We also discuss how non-Markovianity is independent of the generation of entanglement between the system and its environment.
In Section \ref{conclusion}, we summarize our results and formulate some perspectives. Some more technical material is presented in the Appendix.

\section{Definitions and system}
\label{definitions}

\subsection{Measures of non-Markovianity}

Different measures of non-Markovianity have been proposed in the literature, relying on different notions of non-Markovianity. Although these notions are not equivalent, they coincide in many instances~\cite{rivas_quantum_2014,breuer_colloquium_2016,Zheng2011}. In all cases,
non-Markovianity appears as a property of the dynamics, i.e.\ it does not depend on a particular choice of the initial state(s). 
The Rivas-Huelga-Plenio (RHP) measure is based on the divisibility of the dynamical map for the reduced system~\cite{rivas_entanglement_2010}, the Breuer-Laine-Piilo (BLP) measure is devised from information-theoretical considerations~\cite{laine_measure_2010} and the measure introduced in \cite{lorenzo_geometrical_2013} relies on a  geometrical characterization of the dynamics. The RHP measure quantifies the divisibility of the super-operator describing the time evolution of the reduced density matrix. It can be reformulated as a positivity constraint on the rates of the dynamical equation for the density matrix, when this equation can be cast into GKSL form \cite{rivas_quantum_2014}, see Sec.~\ref{compmeas}. The BLP measure is probably the most intuitive: it consists of tracking the time evolution of the trace distance between two initially distinct states of the system. When the trace distance is growing, that may be interpreted as back-flow of information to the system~\cite{laine_measure_2010}, hence a signature of non-Markovianity (see Sec.~\ref{compmeas}), despite some recent qualification of this interpretation \cite{wudarski_petruccione_2016,budini_2018}. A practical limitation of this measure is that it requires an optimization over the two initial states, which becomes prohibitive when studying large systems.

The measure of non-Markovianity introduced in \cite{lorenzo_geometrical_2013} relies on the parametrisation of the system's density matrix by a Bloch vector, see e.g.~\cite{mahler_quantum_1998}. The time evolution is then described by a matrix. The derivative of the determinant of this matrix tells us whether the norm of the Bloch vector is expanding or contracting. Any expansion, i.e.\ when the derivative of the determinant is positive, is defined as a non-Markovian episode in the time evolution. This corresponds to an increase with time of the volume of accessible states. In contrast, for a Markovian dynamics, the volume of accessible states can only decrease with time. This measure is especially well suited for analytical results and will be mainly considered in this work. It will be compared to the two previously introduced measures only in the simplest cases.

\subsection{Spin ensemble with pairwise interaction and local longitudinal field}

We are interested in estimating how the time dynamics of a subset of a system of spins can show non-Markovian features. As our formalism allows us to address a quite general problem, we will first express it in a most general framework. Then our results will be applied to the particular case of a spin-$1/2$ chain. From now on, we set $\hbar=1$.
We consider a set of $N$ spins with spin quantum number $S$ interacting with each other only through pairwise interaction. Moreover each spin is subject to a local longitudinal field. The Hamiltonian describing such a spin ensemble reads
\begin{equation}
  \label{Hgeneral}
  H={-}\sum_{i=1}^N\sum_{j=1}^N \Cij S_i^z S_{j}^z +\sum_{i=1}^N \Gi S_i^z\ ,
\end{equation}
where $S_i^z$ stands for the spin operator in the $z$ direction associated to spin $i$ ($i=1,\ldots,N$), and $\Gi$ is the magnitude of the external field applied on spin $i$. The pairwise correlation matrix $(\Cij)$ is only assumed to be real symmetric and accounts for the geometrical arrangement of the $N$ spins and the range of interaction. Note that at this stage, we do not impose any specific geometry nor boundary conditions. For the sake of simplicity, we restrict ourselves to an external longitudinal field, i.e. in the same direction as the interaction, which allows a fully analytical description of the dynamics. 
The whole set of spins is divided into a subset $\mathcal{S}$ of $p$ spins (labeled hereafter $i=1,\dots, p$ without loss of generality), which defines our system of interest, and the remaining $N-p$ spins ($i=p+1,\dots, N$), which form the environment $\mathcal{E}$. The global system $\mathcal{S}+\mathcal{E}$ is assumed to be isolated, so that it evolves unitarily. If $\rho_{\mathcal{S}+\mathcal{E}}$ denotes its density matrix, it obeys Liouville equation
  \begin{equation}
    \ic \frac{\ud}{\ud t}\rho_{\mathcal{S}+\mathcal{E}}=\left[H, \rho_{\mathcal{S}+\mathcal{E}}\right]\ .
  \end{equation}
The global Hamiltonian (\ref{Hgeneral}) can be written
\begin{equation}
  H=H_\mathcal{S}+H_\mathcal{E}+H_\mathcal{SE}\ , \label{Hsum}
\end{equation}
with
\begin{align}
H_{\cal S}&={-}\sum_{i=1}^p\sum_{j=1}^p \Cij S_i^z S_j^z +\sum_{i=1}^p \Gi S_i^z\ , \label{HqS} \\
H_{\cal E}&= {-}\sum_{i=p+1}^N\sum_{j=p+1}^N \Cij S_i^z S_j^z+\sum_{i=p+1}^N \Gi S_i^z\ ,\label{HqE}\\
H_{\cal SE}&={-}2\sum_{i=1}^p\sum_{j=p+1}^N \Cij S_i^zS_j^z,\label{HqSE}
\end{align}
where $H_{\cal S}$ is the Hamiltonian of the system $\mathcal{S}$ of interest, $H_{\cal E}$ is the Hamiltonian of its environment, and $H_{\cal SE}$ is the interaction Hamiltonian between the system and the environment.
The computational basis states are defined as the common eigenstates of all $S_i^z$ operators ($i=1,\ldots,N$). For convenience, we write these states as
 \begin{equation}
 \left| \bm{s} \bm{\sigma}\right> \equiv \left| \bm{s}\rangle\otimes |\bm{\sigma}\right>=\left| s_1 s_2 \dots s_p\right>\otimes \left| \sigma_{p+1}\sigma_{p+2}\dots \sigma_N\right>\ , \label{comp_basis}
\end{equation}
where $|s_k\rangle$ (resp.\ $|\sigma_k\rangle$) are the eigenstates of $S_k^z$ for $k=1,\ldots,p$ (resp.\ $k=p+1,\ldots,N$) of eigenvalue $s_k\,(\sigma_k) \in\{-S,-S+1,\ldots,S\} $. In particular we use different notation to emphasize the distinction between the system and its environment.
Note that all three Hamiltonians (\ref{HqS}), (\ref{HqE}) and (\ref{HqSE}) are diagonal in the basis (\ref{comp_basis}), and thus pairwise commute.

\section{Non-Markovianity in a spin ensemble with pairwise interaction}
\label{results}

\subsection{Derivation of the main result}
\label{mainderiv}
In this Section, we calculate the reduced density matrix of the system $\mathcal{S}$ at any time $t$ and deduce
from it the witness of non-Markovianity following \cite{lorenzo_geometrical_2013}. 
The time evolution operator of the global system associated to (\ref{Hgeneral}), $U(t)=e^{-\ic (H_\mathcal{S}+H_\mathcal{E}+H_\mathcal{SE}) t}$, acts on the computational basis states as
\begin{equation}
  U(t)\left| \bm{s} \bm{\sigma}\right> =  e^{-\ic [H_\mathcal{S}(\bm{s})+H_\mathcal{SE}(\bm{s},\bm{\sigma})+H_\mathcal{E}(\bm{\sigma})] t}\left| \bm{s} \bm{\sigma}\right>\ ,
\label{diag}
\end{equation}
where
\begin{align}
H_{\cal S}(\bm{s})&={-}\sum_{i=1}^p\sum_{j=1}^p \Cij s_i s_j +\sum_{i=1}^p \Gi s_i\ , \label{HclasS} \\
H_{\cal E}(\bm{\sigma})&={-}\sum_{i=p+1}^N\sum_{j=p+1}^N \Cij \sigma_i \sigma_j+\sum_{i=p+1}^N \Gi \sigma_i\ ,\label{HclasE}\\
H_{\cal SE}(\bm{s},\bm{\sigma})&= {-}2\sum_{i=1}^p\sum_{j=p+1}^N \Cij s_i\sigma_j\label{HclasSE}
\end{align}
are the corresponding scalar Hamiltonians introduced in correspondence to (\ref{HqS})-(\ref{HqSE}) and contain all the physical description of the dynamics. 

We consider a density matrix of the global system
that is initially a product state with respect to the bi-partition $\mathcal{S}+\mathcal{E}$,
\begin{equation}
  \rho_{\mathcal{S}+\mathcal{E}}(0)=\rho_\mathcal{S}(0)\otimes\rho_\mathcal{E}(0)\label{rho_init_sep}\ ,
\end{equation}
In particular, if the initial state of the whole chain is separable it may \emph{not} stay so during the dynamics. It will stay separable only for some prescribed choices of the initial density matrix of both the system and its environment. This important point about possible creation of entanglement during the time evolution, already present within our simple model, will be discussed in more details in Sect.~\ref{correlations} below.
The reduced density matrix of $\mathcal{S}$ at any time $t$ is given by
\begin{equation}
  \label{defrhoS}
  \rho_\mathcal{S}(t)= \tr_\mathcal{E} (\rho_{\mathcal{S}+\mathcal{E}}(t)),\, \mathrm{with}\, \rho_{\mathcal{S}+\mathcal{E}}(t)=e^{-\ic H t} \rho_\mathcal{S+E}(0) e^{\ic H t} ,
\end{equation}
where $\tr_\mathcal{E}$ denotes a partial trace over the environment degrees of freedom. This expression can be explicited as
\begin{equation}
\begin{aligned}
  \rho_\mathcal{S}(t)=&\sum_{\bm{\sigma}} \left< \bm{\sigma}\right| \rho_{\mathcal{S}+\mathcal{E}}(t) \left| \bm{\sigma}\right>\\
  = &  \sum_{\sigma_{p+1}=-S}^S \sum_{\sigma_{p+2}=-S}^S \dots \sum_{\sigma_{N}=-S}^S \left< \bm{\sigma}\right|\rho_{\mathcal{S}+\mathcal{E}}(t) \left| \bm{\sigma}\right> .
\end{aligned}
\label{deftrE}
\end{equation}
Expanding the initial state of the environment in the computational basis as
\begin{equation}
  \rho_\mathcal{E}(0)=\sum_{ \bm{\sigma}', \bm{\sigma}''} a_{ \bm{\sigma}', \bm{\sigma }''} \left|  \bm{\sigma}'\right>\left< \bm{\sigma}''\right|,\label{rhoE_gen}
\end{equation}
the evolved reduced density matrix follows from (\ref{diag}), (\ref{defrhoS}) and (\ref{deftrE})
\begin{equation}
 \left< \bm{s}\right| \rho_\mathcal{S}(t)\left| \bm{s'}\right>={} e^{\ic t [H_\mathcal{S}(\bm{s'})-H_\mathcal{S}(\bm{s})]}  \left< \bm{s}\right| \rho_\mathcal{S}(0)\left| \bm{s'}\right>\A \label{rhoS_t_exact}
\end{equation}
with
\begin{equation}
\A = \sum_{\bm{\sigma}} a_{ \bm{\sigma}, \bm{\sigma}} \,e^{\ic t [H_\mathcal{SE}(\bm{s'},\bm{\sigma})-H_\mathcal{SE}(\bm{s},\bm{\sigma})]}.\label{defA}
\end{equation}
In Eq.~(\ref{defA}), the sum runs over the diagonal elements of the expansion (\ref{rhoE_gen}), which comes from the fact that the Hamiltonian of the environment is diagonal in the computational basis.
Therefore the reduced density matrix of the system $\mathcal{S}$ only depends on the initial populations of the environment in the computational basis. Equations~(\ref{rhoS_t_exact}) and (\ref{defA}) show that the populations of the system are conserved during the dynamics as, for $ \bm{s}= \bm{s'}$, we have $A_{\bm{s},\bm{s}}(t)=1$ for all $t$. This means that the dynamics of the system is purely dephasing.
Using the definition (\ref{HclasSE}) of the interaction Hamiltonian, Eq.~(\ref{defA}) becomes
\begin{equation}
\A = \sum_{\bm{\sigma}} a_{ \bm{\sigma}, \bm{\sigma}} \exp\left[2\ic t \left(
\displaystyle\sum_{j=p+1}^N \sigma_j\sum_{i=1}^p \Cij  (s_i-s'_i) \right) \right].\label{A}
\end{equation}
The next step consists of writing the Bloch vector parametrising the density matrix (\ref{rhoS_t_exact}) in order to compute the determinant of the time evolution operator for the reduced density matrix. This operator is represented by a matrix $M_{\mathcal{S}}(t)$ acting on the Bloch vector, and its determinant is the volume of accessible states. Its exact expression and the calculation of its determinant is a bit lengthy and can be found in Appendix~\ref{Appendix_A}. One eventually gets the closed formula
\begin{equation}
  \label{general_detM}
  \det M_{\mathcal{S}}(t)= \prod_{\bm{s},\bm{s'}} \A\ ,
\end{equation}
where the product over $\bm{s}$ is meant to browse all the eigenstates of the Hamiltonian (\ref{HclasS}), i.e.\ all the $(2S+1)^p$ values of the coordinates of $\bm{s}$ with $s_i=-S,\ldots, S$, and the same for $\bm{s'}$. We find that there is no dependence on the external field. Equation~(\ref{general_detM}) is one of the main results of our paper. 
Following \cite{lorenzo_geometrical_2013}, the dynamics of $\mathcal{S}$ defined by (\ref{defrhoS})
will be non-Markovian whenever 
\begin{equation}
\frac{\ud}{\ud t}\det M_{\mathcal{S}}(t)> 0.
\label{NMwitness}
\end{equation}
This result leads to several remarks.
First, Eqs.~(\ref{A}) and (\ref{general_detM}) show that the couplings between any two spins within the system $\mathcal{S}$ (or the environment $\mathcal{E}$) do not influence the non-Markovianity of $\mathcal{S}$. Instead, non-Markovianity is a feature that only stems from the couplings between $\mathcal{S}$ and $\mathcal{E}$. Second, when the environment is in a computational basis state $\rho_{\mathcal{E}}=|\boldsymbol{\sigma}'\rangle\langle \boldsymbol{\sigma}'|$, the determinant simplifies to $\det M_{\mathcal{S}}(t)=1$ for all times, and the dynamics is Markovian. Last, let us emphasize that the result (\ref{general_detM}) is very general as it is valid for any pairwise interaction strengths $\Cij$ and in particular, for random interactions or for spin glasses \cite{sherrington_solvable_1975}.

\subsection{Application to spin-$1/2$ chains}

Let us exemplify Eq.~(\ref{general_detM}) in the case of $N$ spin-$1/2$. For the sake of simplicity, we consider the environment initially in the maximally mixed state
\begin{equation}
  \rho_\mathcal{E}(0)=\frac{\mathbb{1}_\mathcal{E}}{2^{N-p}}.
  \label{rhoE_mixed}
\end{equation}
Inserting (\ref{rhoE_mixed}) into (\ref{A}), and performing the 
sum over the environment states by descending recursion, we obtain
\begin{equation}
\A =\displaystyle\prod_{j=p+1}^N
\cos\left[\left(\sum_{i=1}^p \Cij(s_i-s'_i) \right)t \right]\label{general_rhoS_t_exact}
\end{equation}
with $s_i,s'_i\in\{-1/2,1/2\}$. We will use this result to determine when one-dimensional spin chains with periodic boundary conditions display non-Markovianity. We are more particularly interested in studying how the range of the interaction can affect the Markovian character of the dynamics of the system $\mathcal{S}$. We will start by investigating the most common case of nearest neighbor interaction. Then we will study the formal case of infinite range where all the spins of the chain interact with each other. Last, we consider a model with power law range, which interpolates between those two situations.

\subsubsection{Ising model with nearest neighbor interaction}

We consider now a spin chain where each spin interacts only with its two nearest neighbors (nn). When comparing with the general form (\ref{Hgeneral}), this amounts to take $\Cij=0$ for $i=j$ and $(|i-j|~\mathrm{mod}~N)>2$, and $\Cij=J$ ($J>0$) for $(|i-j|~\mathrm{mod}~N)=1$. In this case, Eq.~(\ref{general_rhoS_t_exact}) yields
\begin{equation}
   \A=\cos\left[J t\left(s_p-s'_p \right) \right]
\cos\left[J t\left( s_1-s'_1 \right) \right]\ ,\label{QIshrlongh_rhoS_t_exact}
\end{equation}
where it was assumed that the environment contains more than one spin ($N>p+1$).
This explicit expression allows us to evaluate the determinant of the time evolution operator $M_{\mathcal{S}}(t)$ of the Bloch vector given by Eq.~(\ref{general_detM}),
\begin{equation}
  \det M_{\mathcal{S},\mathrm{nn}}(t) =\cos^{2^{2p}}\left( Jt\right),\, N-p\ge 1. \label{detM_QIshrlongh}
\end{equation}
This result indicates that the dynamics of the system is always non-Markovian following the criterion (\ref{NMwitness}), as the derivative of this expression always reaches positive values. Interestingly Eq.~(\ref{detM_QIshrlongh}) does depend neither on the sign of the interaction, nor on the size of the bath. Therefore the system remains non-Markovian in the thermodynamic limit of infinitely large environment ($N\to\infty$).
Another choice of the thermodynamic limit can be taken by choosing a system size which is a finite fraction of the whole chain: $p=rN$. From (\ref{detM_QIshrlongh}) it can be immediately seen that the determinant is zero almost everywhere~\footnote{Except for the set of points where $\cos\left( Jt\right)=\pm 1$, which is of measure zero. So we neglect it for the discussion of non-Markovianity.} so that the dynamics becomes Markovian in this limit.

\subsubsection{Infinite range Ising model}

In this Section, all spins are assumed to be coupled with each other with the same interaction strength, i.e.\ $\Cij=J/N$ ($J>0$) for $i\ne j$ and zero otherwise. In particular, we recover for $p=1$ the case of a single spin coupled uniformly to an environment of spins: this is the celebrated central spin model, which has been extensively studied before, see e.g.~\cite{coish_hyperfine_2004,fischer_correlated_2007,bhattacharya_exact_2017}. 
Note that the Hamiltonian (\ref{HqS}) of the system $\mathcal{S}$ depends on the size of the environment through the interaction constant $J_{ij}=J/N$. This convention is particularly relevant in order to consider the thermodynamic limit as in this case the interaction part of the Hamiltonian follows the same scaling when $N\to\infty$ as the external field part. Evaluating Eq.~(\ref{general_rhoS_t_exact}) and inserting the result into Eq.~(\ref{general_detM}) yields
\begin{equation}
  \det M_{\mathcal{S},\mathrm{\infty}}(t) =\prod_{\bm{s},\bm{s'}} \cos^{N-p}\left[\frac{J t}{N} \sum_{i=1}^p (s_i-s_i')\right].
  \label{QIinfrlongh_detM_anyp_1}
\end{equation}
This expression can be further simplified using a simple combinatorial argument. When varying the spin variables $s_i$'s, each of them being $\pm 1/2$, the sum of them is
\begin{equation}
  \label{sum_si}
  \sum_{i=1}^p s_i=\frac{p-2k}{2}, \quad \binom{p}{k} \textrm{ times}\ , 0\le k \le p.
\end{equation}
The determinant allowing us to estimate the non-Markovianity of the dynamics is then given by
\begin{equation}
 \det M_{\mathcal{S},\mathrm{\infty}}(t) =\prod_{j=0}^p \prod_{k=0}^p \left[\cos\left(\frac{J t}{N} (j-k)\right)\right]^{(N-p)\binom{p}{k}\binom{p}{j}}. \label{QIinfrlongh_detM_anyp}
\end{equation}
In this case again, the witness of non-Markovianity does not depend on the sign of the interaction.
We shall now consider two thermodynamic limits: when the system size is fixed and the environment size goes to infinity, and when the system $\mathcal{S}$ consists of a finite fraction of the whole system $\mathcal{S+E}$, i.e.\ $p=rN$, and $N$ goes to infinity.

The first thermodynamic limit is almost trivial. The product (\ref{QIinfrlongh_detM_anyp}) contains a finite number of factors. One can use for each factor the Taylor expansion
\begin{equation*}
  \cos\left(\frac{J t}{N} (j-k)\right)^{N-p}\simeq \left(1-\frac{(J t)^2(j-k)^2}{2N^2} \right)^{N-p}\ ,
\end{equation*}
to see that each of them will go to $1$ in the limit $N\to\infty$. Eventually one gets
\begin{equation}
 \det M_{\mathcal{S},\mathrm{\infty}}(t)=1\label{detM_QIinfrlongh}\ .
\end{equation}
Following the criterion (\ref{NMwitness}) this means that the system's dynamics is Markovian in this thermodynamic limit. Another way to understand this result is that, in this limit, all the coefficients defined in (\ref{A}) become $\A=1$ so that the system's dynamics (\ref{rhoS_t_exact}) is the same as if it was isolated hence becomes Markovian.

The second thermodynamic limit, which consists of $p=rN$, i.e.\ both the system and its environment have a infinitely growing size, requires a bit more care. 
First, counting each index pair once and doing {the} change of variable $q\equiv k-j$, {Eq.~}(\ref{QIinfrlongh_detM_anyp}) can be rewritten
\begin{equation}
  \det M_{\mathcal{S},\mathrm{\infty}}(t)=\left[\prod_{q=1}^{rN} \cos^2\left(\frac{J t q}{N}\right)^{\sum_{k=q}^{rN}\binom{rN}{k}\binom{rN}{k-q}}\right]^{N(1-r)}\ .\label{detM_QIinfr_thermolim_v0}
\end{equation}
This expression is convenient to see that $\det M_{\mathcal{S},\mathrm{\infty}}(t)$ is a periodic function of $t$ of period $2\pi N/J$.
It reaches the value $1$ when $t$ is an integer multiple of that period. It is enough to restrict ourselves to the behavior during one period. For $0 < t < 2\pi N/J$ at least one factor is smaller than one. As it is raised to a power growing with $N$, it is enough to make the whole product vanish to $0$. This can be more precisely written when $t$ is such that $q Jt/N$ is not a multiple of $\pi$ for any $q$ between $1$ and $rN$.
The exponent of each factor can be simplified by using Chu-Vandermonde identity
\begin{equation}
  \sum_{k=q}^{rN}\binom{rN}{k}\binom{rN}{k-q}=\sum_{k=0}^{rN-q}\binom{rN}{k}\binom{rN}{k+q}=\binom{2 rN}{rN-q}\ .
\end{equation}
Each factor of the product (\ref{detM_QIinfr_thermolim_v0}) is Taylor expanded so that the whole product becomes
\begin{equation*}
  \det M_{\mathcal{S},\mathrm{\infty}}(t)\simeq\left[\prod_{q=1}^{rN} \left(1-\frac{1}{2}\left(\frac{J t q}{N}\right)^2\right)^{\binom{2 rN}{rN-q}}\right]^{2N(1-r)}\ ,
\end{equation*}
which can be rewritten
\begin{equation}
  \label{detM_QIinfr_thermolim_pN}
  \det M_{\mathcal{S},\mathrm{\infty}}(t)\simeq \exp\left[ -\frac{1-r}{N}\left({J t}\right)^2 \sum_{q=1}^{rN} \binom{2rN}{rN-q} q^2 \right] \ .
\end{equation}
As the sum grows at least exponentially when increasing $N$, the determinant converges to $0$ {for all times}, which means that the dynamics is Markovian in this limit.
\subsubsection{Power law range Ising model}

Here a slightly more general model of spin system is investigated, which includes as limiting cases both the previous examples. Consider a one-dimensional chain, where the interaction between any two spins depends on the distance between those spins through a power law (PL). More specifically, the pairwise correlation matrix is chosen as $\Cij=J_N(\alpha)/r_{ij}^\alpha$ ($J_N(\alpha)>0$) for $i\ne j$ and zero otherwise, where $\alpha$ is the parameter ruling the range of the interaction, and $r_{ij}$ denotes the distance between the $i$th and $j$th sites. The interaction strength $J_N(\alpha)$ depends both on $N$ and $\alpha$.
This model is convenient to interpolate between the more common nearest-neighbor interaction ($\alpha\to\infty$) and the infinite range interaction ($\alpha\to 0$). Note that this model for $\alpha=3$ is similar to the RKKY model~\cite{ruderman_indirect_1954,kasuya_electrical_1956,yosida_magnetic_1957}, and has been previously intensively studied in a spin glass perspective, see e.g.~\cite{walker_computer_1980,haussler_interrelations_1992}.
Using Eqs.~(\ref{general_rhoS_t_exact}) and (\ref{general_detM}), the witness for non-Markovianity for the dynamics of $\mathcal{S}$ is obtained by checking the variations of 
\begin{equation}
  \label{QIpowrlongh_detM}
  \det M_{\mathcal{S},\mathrm{PL}}(t)=\prod_{\bm{s},\bm{s'}}
 \prod_{j=p+1}^N
\cos\left[J_N(\alpha) t\left(\sum_{i=1}^p \frac{s_i-s'_i}{r_{ij}^\alpha}  \right) \right].
\end{equation}

Again it is customary to ask whether non-Markovianity survives at the thermodynamic limit of large size. There can be two options for the choice the interaction constant $J_N(\alpha)$: it can be independent of $N$ as for optical atom systems \cite{beguin_direct_2013,barredo_demonstration_2014,saffman_quantum_2010}, or it can scale with $N$ to have a unit mean field temperature, see e.g.~\cite{katzgraber_monte_2003}.
In both cases we can argue qualitatively the same behavior for the non-Markovian character dynamics of the system. Similarly to the previous case of infinite range system, the non-Markovianity witness is a product of periodic functions. The crucial difference is that all the factors display now incommensurable frequencies.   Therefore we predict that, in the large $N$ limit, the whole product should vanish, which is supported by our numerics. In other words the product (32), which is a special case of (22), contains an infinite number of factors. Each of them are raised to a power growing with $N$ so that they become non-zero only for a discrete set of times in the thermodynamic limit. This set is different for each factor so that the whole product vanishes for all time. The situation is different as soon as the support of the interaction is finite. This means that only a finite number of $J_{ij}$ in (22) are non zero. The product now contains a finite number of oscillating factor, hence can generically have piecewise a positive derivative. This is the reason why we conjecture that the dynamics is Markovian at all times whenever the support of the interaction between the system and its environment is infinite, and can become non-Markovian in the case of finitely supported interaction.

\subsection{Influence of the dimension and of the temperature}

\subsubsection{Higher dimensional spin lattice}

It is worth emphasizing that our results~(\ref{rhoS_t_exact}),(\ref{A}) and (\ref{general_detM}) can be applied to other partitions. This is particularly relevant for higher dimensional model. For the sake of illustration we will investigate the case of spins-$1/2$ located on a two-dimensional square lattice interacting via a nearest neighbor interaction, and with periodic boundary conditions.
In order to use our general results, we consider a lattice $\mathcal{S}+\mathcal{E}$ made of $N=M^2$ sites. The systems $\mathcal{S}$ here consists of the $p=q^2$ spins in the square sub-lattice in the upper left corner. Each spin $s_{ix,iy}^z$ is now labeled with two spatial indices $(ix,iy)$, which locates its position along both directions of the lattice. These two indices can be combined in a single index $i$ ranging from $1$ to $N=M^2$ using $i=(ix-1) {M}+(iy-1)+1.$
In order to facilitate the physical interpretation, we will use the 2d indices $(ix,iy)$ in the following discussion. The nearest-neighbor interaction corresponds to the pairwise correlation matrix given by $J_{(ix,iy),(jx,jy)} = J$ ($J>0$) for  $(jx,jy)=(ix,iy+1)$, $(ix,iy-1)$, $(ix+1,iy)$ and $(ix-1,iy)$   and $J_{(ix,iy),(jx,jy)}= 0$ otherwise. In order to ensure periodic boundary conditions, an index taking the value $0$ (resp. $M+1$) corresponds to $M$ (resp. $1$).
The initial state of the environment is, in analogy with the one dimensional case (\ref{rhoE_mixed}),
\begin{equation}
  \rho_\mathcal{E}(0)=\frac{\mathbb{1}_\mathcal{E}}{2^{M^2-q^2}}\ .
  \label{rhoE_mixed2d}
\end{equation}
Inserting (\ref{rhoE_mixed2d}) into (\ref{A}) and using the definition of the pairwise correlation matrix given previously leads to
\begin{align}
  \A &=\prod_{i=1}^q \cos[Jt(s_{1,i}-s_{1,i}')]\prod_{i=1}^q \cos[Jt(s_{q,i}-s_{q,i}')]
  \nonumber\\
  &\prod_{i=1}^q \cos[Jt(s_{i,1}-s_{i,1}')]\prod_{i=1}^q \cos[Jt(s_{i,q}-s_{i,q}')]\ .\label{A_shr2d}
\end{align}
The first two products in~(\ref{A_shr2d}) correspond respectively to the coupling of the first and last rows of spins in $\mathcal{S}$ with the environment. Similarly, the last two products in~(\ref{A_shr2d}) correspond respectively to the coupling of the first and last columns of spins in $\mathcal{S}$ with the environment. Therefore, this shows that only the coupling at the boundary between {the system} and the environment contributes to non-Markovianity. This result is similar to the case of the one-dimensional chains with nearest neighbor interaction previously discussed, see Eq.~(\ref{QIshrlongh_rhoS_t_exact}).
The last step consists of computing the non-Markovianity witness using (\ref{general_detM}). The determinant consists of $2^{2q^2}$ factors in two dimensions. There are exactly $2^{2(q^2-1)}$ factors for which $s_{ix,iy}^z$ and $s_{ix,iy}^{\prime\;z}$ are fixed for a given location $(ix,iy)$. One needs to distinguish between $4(q-2)$ edge sites located at $(ix,iy)\in\{(1,i), (i,q), (q,i),(i,1)\}$ for $2\le i\le q-1$ and $4$ corner sites located at $(1,1)$, $(1,q)$, $(q,q)$ and $(q,1)$. Following (\ref{A_shr2d}) the contribution of a given edge site is
\begin{equation*}
  \left[\prod_{s_{ix,iy}^z=\pm 1/2}\prod_{s_{ix,iy}^{\prime\;z}=\pm 1/2} \cos[Jt(s_{ix,iy}^z-s_{ix,iy}^{\prime\;z})]\right]^{2^{2q^2-2}} \ ,
\end{equation*}
whereas the contribution of any of the four corner sites is
\begin{equation*}
  \left[\prod_{s_{ix,iy}^z=\pm 1/2}\prod_{s_{ix,iy}^{\prime\;z}=\pm 1/2} \cos^2[Jt(s_{ix,iy}^z-s_{ix,iy}^{\prime\;z})]\right]^{2^{2q^2-2}} \ ,
\end{equation*}
Multiplying all those contributions leads to the exact formula for the non-Markovianity witness for a two-dimensional {square} lattice
\begin{equation}
   \det M_{\mathcal{S},\mathrm{nn}}(t) =\left[\cos\left( Jt\right)\right]^{q 2^{2q^2+1}},\, M-q\ge 1 \label{detM_QIshrlongh2d} \ .
\end{equation}
Again it is worth stressing that this formula proves that the dynamics of the sub-lattice will remain non-Markovian for an arbitrary size of the surrounding environment. Conversely, when the size of the system is taken as a finite fraction size of its environment $(q=rM)$, its dynamics becomes Markovian.

\subsubsection{Finite temperature state for the environment}

It is worth noticing that our results can be generalized to account for the effect of the temperature. We will illustrate this for the case of the one-dimensional spin$-1/2$ chain with nearest neighbor interaction, and a homogeneous external field.

\begin{figure}[!ht]
  \begin{center}
    \includegraphics[scale=0.9]{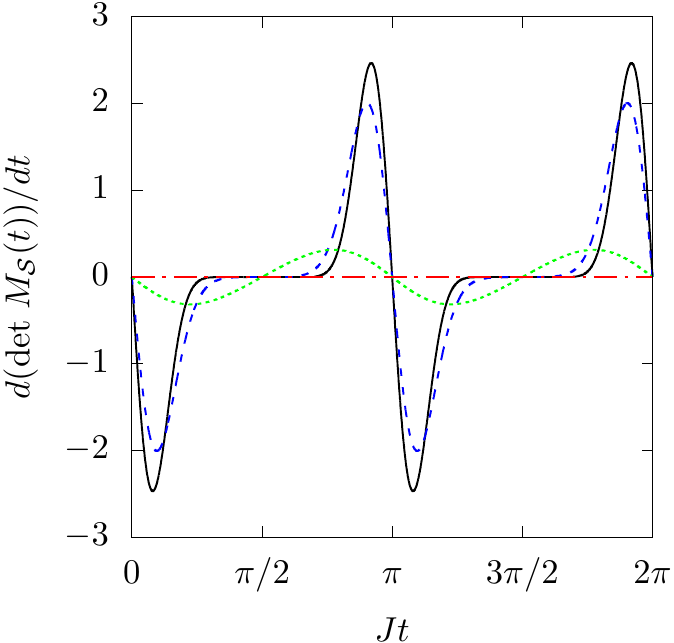}
    \caption{Non-Markovianity witness~(\ref{NMwitness}) of a system of two spin-1/2 with an environment made of $8$ spin-1/2 as a function of time. Here, $\mathcal{S}+\mathcal{E}$ is a chain of $N=10$ spins-1/2 with nearest-neighbor interactions, periodic boundary conditions and $h=J$. The environment is initially in a thermal state~(\ref{rhoE_finiteT}) with $\beta=0$ (black), $\beta =1/J$ (blue dashed), $\beta=3/J$ (green dotted) and $\beta\rightarrow\infty$ (red dot-dashed).}
    \label{NM_vs_T}
  \end{center}
\end{figure}

Start from an initial density matrix for the environment at a given finite temperature $T$:
\begin{equation}
   \rho_\mathcal{E}(0)=\sum_{\bm{\sigma}} \frac{e^{-\beta H_\mathcal{E}(\bm{\sigma})}}{Z}\left| \bm{\sigma}\right>\left<\bm{\sigma}\right|\;,
  \label{rhoE_finiteT}
\end{equation}
where $\beta=1/k_B T$ is the inverse temperature. The Hamiltonian of the environment is, see Eq.~(\ref{HclasE}):
\begin{equation}
  H_\mathcal{E}(\bm{\sigma})=-J\sum_{i=p+1}^{N-1} \sigma_i\sigma_{i+1} + h \sum_{i=p+1}^N \sigma_i\ .
  \label{HqE_open}
\end{equation}
Note that this subchain, defining the environment, obeys open boundary conditions. Last the partition function $Z$ in (\ref{rhoE_finiteT}) is given by:
\begin{equation}
  Z\equiv Z(T,h)=\sum_{\bm{\sigma}} e^{-\beta H_\mathcal{E}(\bm{\sigma})}
\end{equation}
As detailed in \ref{mainderiv} the way to assess the non-Markovian character of the dynamics will be achieved in two steps. First the coefficients $\A$ as defined in Eq.~(\ref{defA}) are computed. Then the determinant (\ref{general_detM}) and its first derivative are evaluated numerically. This is illustrated  in Fig.~\ref{NM_vs_T}, which shows the non-Markovianity witness for different temperatures of the environment. Notice that, similarly to (\ref{detM_QIshrlongh}), the determinant is a periodic function of the time $t$ with period $2\pi/J$. Hence it is plotted only over one period.
It can be seen that the dynamics is non Markovian for any non vanishing temperature. Note that, for $T=0$, the initial density matrix of the environment is $|\boldsymbol{\sigma_0}\rangle\langle \boldsymbol{\sigma_0}|$ in the computational basis, where $\left|\boldsymbol{\sigma_0}\right>$ is the ground state of the Hamiltonian. As mentioned earlier after Eq.~(\ref{NMwitness}), this leads trivially to a Markovian dynamics for the system.

\subsection{Comparison with other non-Markovianity measures for systems of $p=1$ spin}
\label{compmeas}

Our results can be used in order to compare different measures of non-Markovianity \cite{neto_inequivalence_2016}.
For the sake of illustration, let us consider here the special case of a system consisting of one spin-$1/2$ ($p=1$). 
Using~(\ref{general_detM}) and (\ref{general_rhoS_t_exact}) valid for the environment initially in the maximally mixed state, the determinant of the evolution operator is
\begin{equation}
  \det M_{\mathcal{S}}(t)=A(t)^2\ ,
\end{equation}
with 
\begin{equation}
  \label{def_f0}
  A(t)=\prod_{j=2}^N \cos\left(J_{1j} t \right).
\end{equation}
such that, following the criterion (\ref{NMwitness}), the dynamics is non-Markovian whenever
\begin{equation}
  \label{NM_p1}
  A(t)A'(t)>0\ .
\end{equation}
In order to evaluate other witnesses of non-Markovianity, we write explicitly  the reduced density matrix of $\mathcal{S}$ at any time $t>0$. Eq.~(\ref{rhoS_t_exact}) yields
\begin{equation}
  \label{general_rhoS_t_exact_p1}
 \rho_\mathcal{S}(t)=\left(
\begin{array}{cc}
  \rho_{11} & \rho_{12}  A(t)\, e^{-\ic h_1 t} \\
 \rho_{21}  A(t)\, e^{\ic h_1 t} & \rho_{22}
\end{array}\right),
\end{equation}
where $\rho_{ij}$ ($i,j=1,2$) are the coefficients of the initial density matrix of $\mathcal{S}$ at $t=0$.
One can get the corresponding Kraus representation (see e.g.~\cite{andersson_finding_2007}) and deduce from it the master equation for the reduced density matrix~\cite{hall_canonical_2014,bhattacharya_exact_2017}
\begin{equation}
  \frac{\ud}{\ud t} \rho_\mathcal{S}(t)=-\ic\left[H^{\rm eff}_\mathcal{S},\rho_\mathcal{S}(t)\right]
  +\Gamma_z(t) \Big( \sigma^z \rho_\mathcal{S}(t)\sigma^z-\rho_\mathcal{S}(t)\Big)
\label{master_eq_p1},
\end{equation}
with the effective Hamiltonian $H^{\rm eff}_\mathcal{S}=h_1 \sigma^z/2$ where $\sigma^z$ stands for the usual Pauli matrix. 
This master equation models a pure dephasing channel with a time-dependent rate
\begin{equation}
  \Gamma_z(t)=-\frac{A'(t)}{2A(t)}. \label{Gammaz_p1}
\end{equation}
The master equation~(\ref{master_eq_p1}), of the GSKL form, can be used to evaluate divisibility criterion, as it can be expressed as a sign constraint on the rate in the master equation. The RHP measure detects a non-Markovian behavior when the rate in the master equation becomes negative \cite{rivas_quantum_2014}. Due to the explicit expression (\ref{Gammaz_p1}) the dynamics will be non-Markovian if $-A'(t)/A(t)<0$, which trivially agrees with our witness~(\ref{NM_p1}).
Knowing the exact expression~(\ref{general_rhoS_t_exact_p1}) of $\rho_\mathcal{S}(t)$ enables one also to compute BLP distance measure of non-Markovianity~\cite{laine_measure_2010}. The trace distance between two arbitrary states $\rho^a_0$ and $\rho^b_{0}$
is given by
\begin{align}
D(\rho_a(t),\rho_b(t))&=\tr \left(\sqrt{(\rho_a(t)-\rho_b(t))(\rho_a(t)-\rho_b(t))^\dagger}\right)\nonumber\\ 
&=\sqrt{(\rho_{11}^a-\rho_{11}^b)^2+A(t)^2|\rho_{12}^a-\rho_{12}^b|^2}.
\label{BLP_p1}
\end{align}
The system is said to be non-Markovian according to the BLP measure whenever
\begin{equation}
\frac{\ud}{\ud t}D(\rho_a(t),\rho_b(t))=\frac{|\rho_{12}^a-\rho_{12}^b|^2 A(t) A'(t)}{\sqrt{(\rho_{11}^a-\rho_{11}^b)^2+A(t)^2|\rho_{12}^a-\rho_{12}^b|^2}}
\end{equation}
is strictly positive. As here, $0\leq A(t)^2\leq 1 $, and, for any density operator of a two-level system, we have $|\rho_{12}|\leq \rho_{11}\leq 1$ and $|\rho_{12}|\leq 1/2$, see e.g. \cite{HornJohnson}, the maximum of this expression is reached for $\rho_{11}^a=\rho_{11}^b$ and $\rho_{12}^a=-\rho_{12}^b=1/2$. This condition for non-Markovianity is satisfied whenever $A(t) A'(t)>0$ which agrees again with (\ref{NM_p1}).

\section{Entanglement and non-Markovianity}
\label{correlations}

The aim of this Section is to investigate the relation between the non-Markovianity of the system $\mathcal{S}$ and the generation of entanglement with the environment $\mathcal{E}$. Let us remind that we consider an initial state without system-environment entanglement of the form (\ref{rho_init_sep}).

First, let us show that the dynamics of $\mathcal{S}$ can display non-Markovianity, according to the witness~\eqref{NMwitness}, without generating any entanglement with the environment. For this purpose, we consider an initial separable state of the system and the environment as in (\ref{rho_init_sep}), the initial density matrix of the latter being a classical mixture of computational basis state
\begin{equation}
  \rho_\mathcal{E}(0)=\sum_{  \bm{\sigma}'} a_{ \bm{\sigma}', \bm{\sigma}'} \left|   \bm{\sigma}'\right>\left< \bm{\sigma}'\right|.
  \label{rhoE_mixture}
\end{equation}
According to our previous analysis, the system's non-Markovianity is given in this case by Eqs.~(\ref{general_detM})-(\ref{NMwitness}). Writing the initial state of $\mathcal{S}$ as 
\begin{equation}
  \rho_\mathcal{S}(0)=\sum_{ \bm{s},\bm{s}'} r_{ \bm{s}, \bm{s}'} \left|  \bm{s}\right>\left< \bm{s}'\right|
\end{equation}
and using Eqs.~(\ref{diag}),~(\ref{HclasS}),~(\ref{HclasE}), and~(\ref{HclasSE}), we obtain
\begin{equation}
 \rho_{\mathcal{S}+\mathcal{E}}(t) = \sum_{\bm{\sigma}} a_{ \bm{\sigma}, \bm{\sigma}}\left(\rho_{\mathcal{S}|\bm{\sigma}}(t)\otimes\left| \bm{\sigma}\right>\left< \bm{\sigma}\right|\right)
 \label{rhoSE_mixture}
\end{equation}
with the conditional state of the system
\begin{equation}
\begin{aligned}
\rho_{\mathcal{S}|\bm{\sigma}}(t)=\sum_{\bm{s},\bm{s}'}  &e^{\ic t [H_\mathcal{S}(\bm{s}')-H_\mathcal{S}(\bm{s})+H_\mathcal{SE}(\bm{s}',\bm{\sigma})-H_\mathcal{SE}(\bm{s},\bm{\sigma})]}\\ 
&\times r_{ \bm{s}, \bm{s}'} \left|  \bm{s}\right>\left< \bm{s}'\right|.
\end{aligned}
\end{equation}
Therefore, we see that the global system $\mathcal{S}+\mathcal{E}$ stays in a separable state at all times as shown by Eq.~(\ref{rhoSE_mixture}), independently of the non-Markovianity of $\mathcal{S}$.
Moreover, the state~\eqref{rhoSE_mixture} has, by definition, zero discord with respect to the environment \cite{Olliver2001,Henderson2001}.
Note that, similarly, if the system $\mathcal{S}$ starts in a classical mixture of computational basis states, the global system $\mathcal{S}+\mathcal{E}$ stays in a separable state at all times independently of the initial state of the environment. 
This result is in agreement with previous works on qubit-environment entanglement generation during pure dephasing dynamics~\cite{Roszak2015,Roszak2018}.

\begin{figure}[ht]
\begin{center}
\includegraphics[scale=0.9]{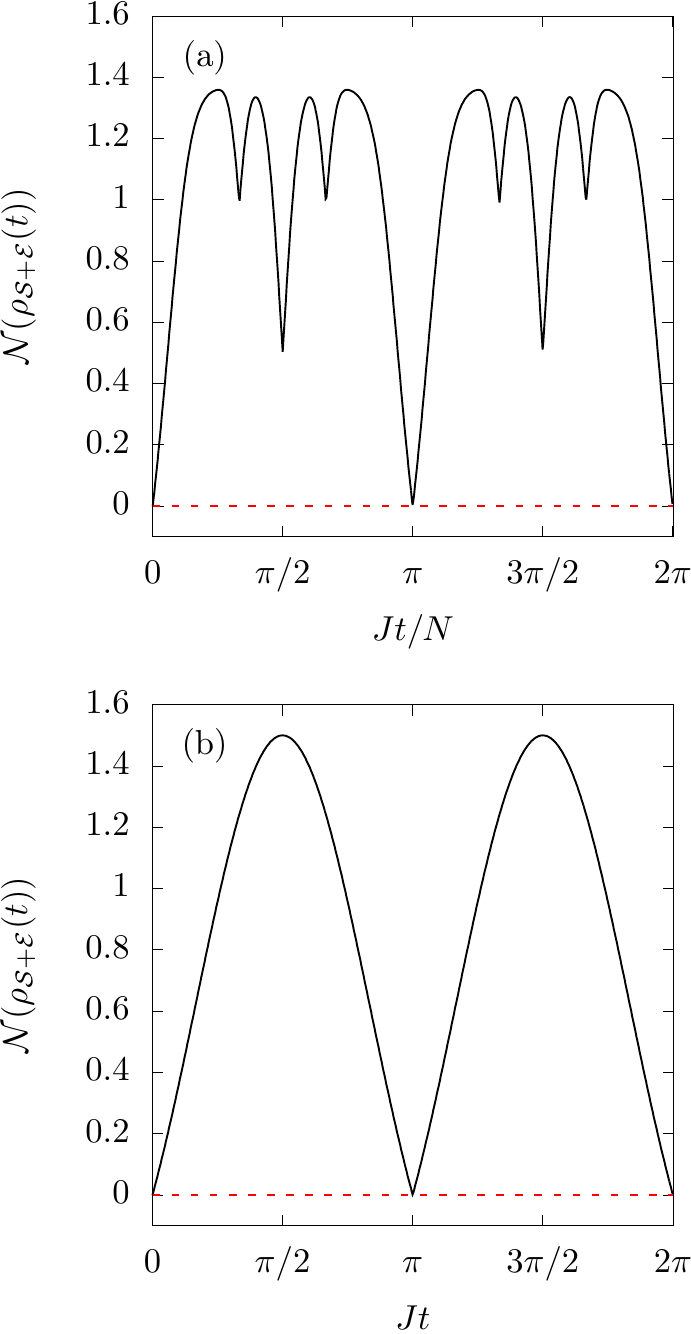}
\caption{Negativity $\mathcal{N}$ between the system $\mathcal{S}$ made of $p=3$ spin-$1/2$ and the environment $\mathcal{E}$ made of $7$ spin-$1/2$, as a function of time for (a) infinite range interactions, and (b) nearest-neighbor interactions with periodic boundary conditions and $h_i=0$ for all $i=1,\dots,10$. The system $\mathcal{S}$ is initially in the pure state $|\psi_\mathcal{S}(0)\rangle=\sum_{\bm{s}}|\bm{s}\rangle/{2^{p/2}}$. The black solid curves correspond to the environment initially in the state $|\psi_\mathcal{E}(0)\rangle=\sum_{\bm{\sigma}}|\bm{\sigma}\rangle/{2^{7/2}}$, and the red dashed curves to the environment initially in a classical mixture of computational basis state~\eqref{rhoE_mixture}.}
\label{fig_neg}
\end{center}
\end{figure}

Let us now show that the system and its environment can get entangled during the dynamics, when the initial state of the environment $\rho_\mathcal{E}(0)$ has non-vanishing coherences $a_{ \bm{\sigma}', \bm{\sigma}''}$ in the computational basis. As an illustration, we consider a chain of $N=10$ spin-$1/2$ with infinite range or nearest neighbors interaction {and various sizes of the system $\mathcal{S}$}. The presence of entanglement between $\mathcal{S}$ and $\mathcal{E}$ is assessed using the negativity
\begin{equation}
\mathcal{N}\left(\rho_{\mathcal{S}+\mathcal{E}}(t)\right)=\frac{||\rho_{\mathcal{S}+\mathcal{E}}^{T_\mathcal{S}}(t)||_1-1}{2}
\end{equation}
where $||\rho||_1=\mathrm{Tr}(\sqrt{\rho\rho^\dagger})$ and $\rho_{\mathcal{S}+\mathcal{E}}^{T_\mathcal{S}}(t)$ is the partial transpose of $\rho_{\mathcal{S}+\mathcal{E}}(t)$ with respect to $\mathcal{S}$. The Peres-Horodecki negativity criterion~\cite{Peres1996,Horodecki1996} states that whenever the negativity is non-zero, the bipartite system $\mathcal{S}+\mathcal{E}$ is entangled. This criterion is necessary and sufficient in the case of two spin-$1/2$ and two spin-$1$. For higher dimensional systems, all separable states have zero negativity but there also exist entangled states with zero negativity. Figure~\ref{fig_neg} illustrates that $\mathcal{N}(\rho_{\mathcal{S}+\mathcal{E}}(t))$ oscillates as a function of time for a system $\mathcal{S}$ made of $p=3$ spins and a given initial state (\ref{rho_init_sep}). Numerical simulations showed that whenever the coherences of the initial density matrix of the environment are non-vanishing, the dynamics typically generates entanglement between the system and its environment.

We have shown in Sec.~\ref{mainderiv} that, for any given separable global state of the form (\ref{rho_init_sep}), the non-Markovianity of the system is independent of the coherences of the initial density matrix of the environment. The reason is that the reduced dynamics of $\mathcal{S}$ given by Eqs.~\eqref{rhoS_t_exact} and~\eqref{A} is independent of the off-diagonal elements of $\rho_\mathcal{E}(0)$.
Yet, having non-zero coherences will lead to the generation of entanglement between the system and its environment, see Fig.~\ref{fig_neg}, whereas the initial density matrix of $\mathcal{E}$ with the same populations and no coherence will lead to a separable dynamics, see (\ref{rhoSE_mixture}). As a consequence, we claim that, for our model, the non-Markovianity is independent of the generation of entanglement between the system and its environment.

\begin{figure}[ht]
\begin{center}
\includegraphics[scale=0.9]{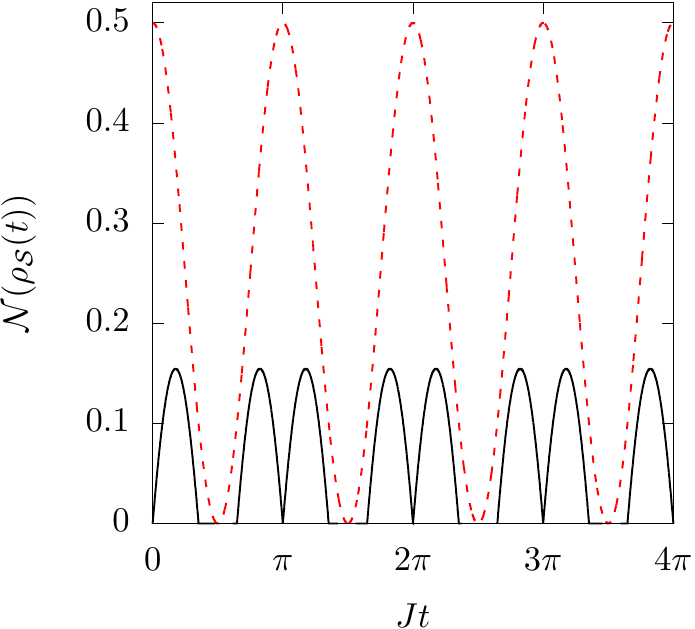}
\caption{Negativity $\mathcal{N}$ between two spin-$1/2$ that define the system $\mathcal{S}$ as a function of time when $\mathcal{S}+\mathcal{E}$ is a chain of $N$ spin-$1/2$ with nearest-neighbor interactions and periodic boundary conditions and $h_i=0$ for all $i=1,\dots,N$. This results is valid for any size of the environment as soon as $N\geq 4$.
The environment is initially in the maximally mixed state~(\ref{rhoE_mixed}). The black curve corresponds to the system $\mathcal{S}$ initially in the separable pure state $\left(|-\tfrac{1}{2}\rangle +|\tfrac{1}{2}\rangle\right)\otimes\left(|-\tfrac{1}{2}\rangle +|\tfrac{1}{2}\rangle\right)/2$ while the red dashed curve corresponds to $\mathcal{S}$ initially in the entangled pure state $\left(|-\tfrac{1}{2}\rangle \otimes |-\tfrac{1}{2}\rangle +|\tfrac{1}{2}\rangle \otimes |\tfrac{1}{2}\rangle\right)/\sqrt{2}$. 
}
\label{fig_neg_S}
\end{center}
\end{figure}

Last, although our model is sufficiently simple to allow for analytical calculations, it is interesting to note that spins within the system $\mathcal{S}$ undergoing non-Markovian dynamics can display non-trivial entanglement dynamics as illustrated in Fig.~\ref{fig_neg_S}. In particular, when $\rho_{\mathcal{S}+\mathcal{E}}(0)$ is a fully separable $N$-spin state, we observe that the system can display sudden-death and revival of entanglement~\cite{bellomo_non-markovian_2007}, whereas the environment stays at all times in a separable state.

\section{Conclusion}
\label{conclusion}

In this paper, we investigated analytically the quantum non-Markovianity of a spin ensemble ($\mathcal{S}$) undergoing a pure dephasing dynamics arising from the unitary evolution of a larger spin ensemble ($\mathcal{S}+\mathcal{E}$) governed by the Hamiltonian (\ref{Hgeneral}). One of our main results is given by Eqs.~(\ref{A})--(\ref{general_detM}) that apply to spin ensembles of arbitrary size and spin quantum number and allows us to determine analytically whether the dynamics is Markovian or not. For a spin-$1/2$ ensemble $\mathcal{S}+\mathcal{E}$ of finite size, we found out that, when the environment $\mathcal{E}$ is initially in the maximally mixed state, the dynamics of $\mathcal{S}$ is always non-Markovian. We also obtained analytical results in the thermodynamic limit for one-dimensional spin chains. In the limit of infinite size of the environment with fixed size of the system, the quantum dynamics of the system stays non-Markovian for nearest-neighbor interactions whereas it becomes Markovian for infinite range interactions, see Eq.~(\ref{detM_QIshrlongh}) vs Eq.~(\ref{detM_QIinfrlongh}). In the limit of infinite size of the environment with the size of the system being a fixed fraction of the ensemble $\mathcal{S}+\mathcal{E}$, the quantum dynamics of the system becomes Markovian both for nearest-neighbor and infinite range interactions. In these limits, we found out that Markovianity can appear when (i) the non-Markovian episodes are separated by a period whose value goes to infinity (cases studied with infinite range interaction), and (ii) the non-Markovian episodes occur for a duration shrinking to zero (cases studied with infinite range  or nearest-neighbor interaction with a system size being a fixed fraction of $N$). We gave another application of our results to two-dimensional square spin lattice. We also showed that, for our system, non-Markovianity does not stem from the generation of entanglement with the environment. Although these observations are specific to our system, they raise the more general question of the relationship between non-Markovianity and system-environment correlations. Natural extensions of this work include the study of dynamics {more general than purely dephasing} or non-integrable dynamics \cite{ignacio_classical_chaos,davalos_quantum_2017}, e.g.\ in the presence of transverse field. Experimental realizations of the system studied in this work could be realized with cold atoms in optical lattices, see e.g.~\cite{navarrete-benlloch_simulating_2011}.

\begin{acknowledgments}
This work was supported by  the  ARC  grant QUANDROPS 12/17-02.
\end{acknowledgments}

\appendix

\section{Determinant of the time evolution operator for the reduced dynamics}\label{Appendix_A}

We start by explaining how to write the time evolution operator of the Bloch vector when the coefficients of the density matrix are explicitly known.
It will be illustrated for the system considered in the main part of the paper: $p$ spin-$S$ interacting via a pairwise interaction, see e.g.~(\ref{HqS}). In particular the dimension of the Hilbert space of the system under consideration is $D=(2S+1)^p$. The Bloch parametrisation for a density matrix $\rho$ of size $D\times D$ (see e.g.~\cite{mahler_quantum_1998}) consists of re-arranging the $D^2$ entries of the density matrix into a vector, called the Bloch vector. The coordinates $r_j$ of the Bloch vector are called the Bloch parameters. They are divided into two sets: one set containing $D(D-1)$ real Bloch coordinates to parametrise the off diagonal elements $\rho_{ij}$ $(i\ne j)$ of the density matrix. They can be grouped in pairs, for the real and the imaginary part respectively. More precisely one can define
\begin{align}
 \hspace{-1cm}r_1=\re(\rho_{12}),\;& r_2=\im(\rho_{12})\nonumber\\
\hspace{-1cm}r_3=\re(\rho_{13}),\;& r_4=\im(\rho_{13})\nonumber\\
  \vdots & \vdots\nonumber\\
  r_{2(D-1)+1}=\re(\rho_{21}),\;&\ r_{2(D-1)+2}=\im(\rho_{21})\nonumber\\
  r_{2(D-1)+3}=\re(\rho_{23}),\;&\ r_{2(D-1)+4}=\im(\rho_{23})\nonumber\\
  \vdots & \vdots\nonumber\\
\hspace{-1cm}r_{D(D-1)-1}=\re(\rho_{D-1\; D}),\; & r_{D(D-1)}=\im(\rho_{D-1\; D})\nonumber
\end{align}
The second set of the $D^2$ Bloch coordinates are formed by $D$ linear combinations of the diagonal elements of the matrix.
\begin{equation*}
  r_{D(D-1)+l}=\sqrt\frac{2}{l(l+1)}\left(\sum_{k=1}^{l} \rho_{kk}-l\rho_{l+1\; l+1}\right)\ ,
\end{equation*}
for $1\le l \le D-1$. The last remaining coefficient is chosen by convention to be
\begin{equation*}
  r_{D^2}=\sum_{k=1}^{D} \rho_{kk}\ ,
\end{equation*}
so that it is unity for a density matrix.
If the $D^2-$dimensional Bloch vector corresponding to the matrix at time $t$ is denoted by $\mathbf{r}(t)$, one can define its time evolution operator $M_{\mathcal{S}}(t)$ through
\begin{equation}
  \mathbf{r}(t)=M_{\mathcal{S}}(t)\,\mathbf{r}(0). \label{prop_vBloch}
\end{equation}
It can be shown that the operator $M_{\mathcal{S}}(t)$ is linear, hence can be represented by a $D^2\times D^2$ matrix.

The explicit expression (\ref{rhoS_t_exact}) allows for a direct evaluation of the coefficients of the matrix representing $M_{\mathcal{S}}(t)$. As the diagonal elements of the density matrix are unchanged, the evolution operator boils down to the identity in the subspace spanned by the second set of Bloch coordinates, as defined above. For the first set, it can be seen directly from (\ref{rhoS_t_exact}) that each pair of Bloch coordinates $(r_{2j-1},r_{2j})$ for $1\le j \le D(D-1)/2$ follow a rotation, expressed by the time dependent phase, and a dilatation expressed by the factor $\A$
\begin{equation}
  \left(
    \begin{array}{c}
      r_{2j-1}(t) \\
      r_{2j}(t)
    \end{array}\right)= \mathcal{O}_j
  \left(
    \begin{array}{c}
      r_{2j-1}(0) \\
      r_{2j}(0)
    \end{array}\right)
\end{equation}
with
\begin{equation}
  \mathcal{O}_j=\left(
    \begin{array}{cc}
      \A \cos \theta_{\bm{s},\bm{s'}}t & \A\sin \theta_{\bm{s},\bm{s'}}t \\
      -\A\sin \theta_{\bm{s},\bm{s'}}t & \A \cos \theta_{\bm{s},\bm{s'}}t
    \end{array}\right)\ ,
\end{equation}
where the notation $\theta_{{\bf s},{\bf s'}}=H_\mathcal{S}(\bm{s'})-H_\mathcal{S}(\bm{s})$ was introduced for the sake of brevity.
In other words the matrix $M_{\mathcal{S}}(t)$ in (\ref{prop_vBloch}) can be written in a block structure for the first set of Bloch coordinates
\begin{equation}
  M_{\mathcal{S}}(t)=\left(
\begin{array}{cccc}
  \mathcal{O}_1 & 0 & \dots & 0\\
  0 & \mathcal{O}_2 & \dots & 0\\
  \vdots & \vdots & \ddots & \vdots \\
  0& 0 & \dots &  \mathcal{O}_{D(D-1)/2}
\end{array}\right)\ ,
\end{equation}
and its determinant is directly given by
\begin{equation}
  \det M_{\mathcal{S}}(t)=\prod_{j=1}^{D(D-1)/2} \det \mathcal{O}_j=\prod_{\bm{s},\bm{s'}} \A \ ,
\end{equation}
which is exactly (\ref{general_detM}).

\bibliographystyle{ieeetr}

\end{document}